\documentclass[conference,letterpaper]{IEEEtran}

\usepackage[utf8]{inputenc} 
\usepackage[T1]{fontenc}
\usepackage{url}
\usepackage{ifthen}
\usepackage{cite}
\usepackage[cmex10]{amsmath}

\newcommand{\dd}{\textnormal{d}}
\usepackage{amssymb} 
\usepackage[center]{caption} 
\usepackage{tikz}
\usetikzlibrary{arrows.meta}
\allowdisplaybreaks

\usepackage[amsmath]{ntheorem}
\usepackage{dsfont}
\usepackage{bbm}
\usepackage{mleftright}
\usepackage{mathtools}
\mleftright

\usepackage{hyperref}
\usepackage{cleveref}
\hypersetup{
  colorlinks=true, 
  linktoc=all,     
  linkcolor=black,  
  citecolor=black
}

\theoremstyle{plain}
\theoremheaderfont{\normalfont\itshape}
\theoremseparator{:}
\newtheorem{theorem}{Theorem}
\newtheorem{lemma}{Lemma}

\newtheorem{proposition}{Proposition}

\providecommand{\abs}[1]{\ensuremath{\left\lvert #1 \right\rvert}}
\providecommand{\norm}[1]{\ensuremath{\left\Vert #1 \right\Vert}}

\begin{document}
\title{Covert Capacity of AWGN Channels\\ under Average Error Probability} 

\author{%
  \IEEEauthorblockN{Cécile Bouette\IEEEauthorrefmark{1}, Laura Luzzi\IEEEauthorrefmark{1}\IEEEauthorrefmark{3} and Matthieu Bloch\IEEEauthorrefmark{2}} 
  \IEEEauthorblockA{\IEEEauthorrefmark{1}ETIS (UMR 8051, CY Cergy Paris Université, ENSEA, CNRS), Email: \{laura.luzzi, cecile.bouette\}@ensea.fr}
  \IEEEauthorblockA{\IEEEauthorrefmark{3}\'Equipe COSMIQ, Inria de Paris, 
    Email: laura.luzzi@inria.fr}
  \IEEEauthorblockA{\IEEEauthorrefmark{2}School of Electrical and Computer Engineering, Georgia Institute of Technology, Atlanta GA 30332, USA,\\ Email: matthieu.bloch@ece.gatech.edu}
}
\maketitle

\begin{abstract} We derive upper and lower bounds for the covert capacity of Additive White Gaussian Noise channels when measuring reliability in terms of the average error probability and covertness in terms of Kullback-Leibler divergence. This characterization confirms the absence of strong converse for this setting in both the reliability and covertness parameters. The crux of our approach is to analyze a codebook of BPSK-modulated codewords carefully augmented with ``all-zero'' codewords.
\end{abstract}

\section{Introduction}
\label{sec:introduction}
Covert communication~\cite{fundamental_covertness,bloch_resolvability,bash_article} addresses the challenge of enabling reliable communication between legitimate parties while avoiding detection by adversaries. Unlike secrecy, which focuses on concealing the content of a transmitted message, covertness aims to hide the very existence of the transmission~\cite{hou2014effective}. This property is essential in scenarios where revealing sensitive attributes, such as who is communicating, when, and from where, must be avoided.

For discrete memoryless channels (DMCs) and additive white Gaussian noise (AWGN) channels, the standard capacity under a covertness constraint is zero because the maximum amount of reliably and covertly transmitted information scales with the square root of the total number of channel uses~\cite{bash_article}. This scaling phenomenon is known as the \emph{square root law} of covert communications. Despite this limitation, the covert capacity can be defined as the constant associated with the square-root scaling. When the decoding error probability vanishes with increasing the block length, covert capacity has been characterized for a broad range of channels, including classical DMCs~\cite{fundamental_covertness,bloch_resolvability,tahmasbi2018first,MIMO_covertness}, AWGNs~\cite{fundamental_covertness,wang2021covert,wang19,zhang2019undetectable,xinchun2024second}, and various additive memoryless channels~\cite{bouette_journal_covert_over_additive_channels}, as well as classical-quantum channels, both with and without entanglement assistance~\cite{Wang2016c,Bullock2020,Gagatsos2020Covert,Wang2024Resource}. Extensions to multi-user settings also exist for multiple-access~\cite{Arumugam2018a,Bounhar2023Mixing}, interference~\cite{Cho2021Treating} and broadcast~\cite{Tan2019} channels, often yielding surprising simplifications compared to the non-covert setting.

However, the notion of covert capacity involves several subtleties. First, it depends on the choice of the covertness metric and the parameter $\delta > 0$ that defines the covertness level~\cite{fundamental_covertness,tahmasbi2018first}. Second, and particularly relevant to this work, the covert capacity may or may not depend on the parameter $\epsilon>0$ that measures reliability. Specifically, under a \emph{maximal} error probability constraint, the covert capacity is independent of $\epsilon$~\cite{tahmasbi2018first}. In contrast, under an \emph{average} error probability constraint, a strong converse does not hold, and the covert capacity depends on $\epsilon$. This behavior arises from the ability to include a controlled number of ``all-zero'' undetectable codewords in the codebook, which increases the codebook size without significantly affecting the covertness constraint or the average error probability~\cite[Appendix A]{tahmasbi2018first}. This is related to the absence of strong converse channel coding over AWGN under average power and average error probability~\cite[Theorem 77]{polyanskiythesis}.

Our main contribution is to study the covert capacity for a positive average error probability $\epsilon$ and derive upper and lower bounds for the covert capacity of AWGN channels. We demonstrate that allowing a small positive error probability $\epsilon$ enables the transmission of additional covert information. The crux of our approach is a careful random coding argument for BPSK-coded messages in which the code is augmented with all-zero codewords, along with an associated converse proof. 

The remainder  of the paper is organized as follows. In Section~\ref{sec:notation}, we review the notation used throughout. In Section~\ref{sec:setup}, we introduce our system model and associated results. Proofs are delegated to Sections~\ref{sec:converse} and~\ref{sec:achievability} to streamline the presentation.

\section{Notation}
\label{sec:notation}
Vectors of length $n$ are denoted with a superscript $n$. We use upper-case letters to denote (real-valued) random variables and lower-case letters to denote their realizations. A length-$n$ random vector $(X_1,\ldots,X_n)$ is denoted $X^n$. We use $P_X$ to denote the distribution of the random variable $X$. When it exists, the probability density function (PDF) corresponding to $P_X$ is denoted $p_X$.  We denote the product of measures by $\otimes$.
and the mutual information between $X$ and $Y$ is denoted $I(X;Y)$~\cite{coverthomas06}; all of these are measured in nats.

Let $P_1$ and $P_2$ be two distributions on the same measurable space $\mathcal{X}$ such that $P_1$ is absolutely continuous with respect to $P_2$, the Kullback-Leibler (KL) divergence between $P_1$ and $P_2$ is 
\begin{IEEEeqnarray}{rCl}
  \label{def_kl_divergence}
  {D}(P_1\|P_2)&=&\int_{\mathcal{X}}\ln\left(\frac{\dd P_1}{\dd P_2}\right) \dd P_1.
\end{IEEEeqnarray}
A random variable $Z$ following a Gaussian distribution with mean $\mu$ and standard deviation $\sigma>0$ is denoted $Z\sim\mathcal{N}(0,\sigma)$.

\section{System Model and Main Results}
\label{sec:setup}

\begin{figure}[tbp]
  \begin{center}
    \begin{tikzpicture}[
  nodetype1/.style={
    rectangle,
    minimum width=0.5cm,
    minimum height=0.7cm,
    draw=black,
    font=\normalsize
  },
  nodetype2/.style={
    rectangle,
    minimum width=0.4cm,
    minimum height=0.6cm,
    draw=black,
    font=\normalsize
  },
  tip2/.style={-{Stealth[length=0.6mm, width=0.5mm]}}
  ]
  \matrix[row sep=0.3cm, column sep=0.3cm, ampersand replacement=\&]{
    \& \& (invisible) \&\&
    \node (Key) [draw, nodetype1, text width=1.2cm, text centered]  {\text{key} $K$}; \&\& \\
    \node (Alice) {\text{transmitter}};  \& \& \node (encoder) [nodetype2]   {$f$}; \&
    \node (X){}; \&
    \node (W) [draw, circle, text centered,inner sep=0pt]  {\text{\Large$+$}}; \&
    \node (Y){};
    \&
    \node (decoder) [nodetype2] {$g$}; \&
    \node (Bob) {\text{receiver}};\\
    \& \& \&
    \& 
    \node (Z) {$Z^n$};
    \& \& \&
    \node (Eve) {\text{eavesdropper}}; \& \\};
  
  \draw[->] (Alice) edge[tip2] node [above] {$W$} (encoder) ;
  \draw[->] (encoder) edge[tip2] node [above] (X) {} (W) ;
  \draw[-]  (encoder.east) -- node [above] (X1) {$X^n$} (W.west) ;
  \draw[arrows = {-Latex[length=1pt]}] (W) edge[tip2] node [above] {$Y^n$} (decoder) ;
  \draw[->] (decoder) edge[tip2] node [above] {$\hat{W}$} (Bob) ;
  \draw[-{Stealth[length=0.5mm, width=0.5mm]}] (Y.center)  |- node [above] {} (Eve) ;
  \draw[-{Stealth[length=0.5mm, width=0.5mm]}] (Z.north)  -- node [above] {} (W) ;
  \draw[{Stealth[length=0.5mm, width=0.5mm]}-] (encoder) |-  (Key) ;
  \draw[-{Stealth[length=0.5mm, width=0.5mm]}] (Key) -| (decoder) ;
\end{tikzpicture}
    \caption{Covert communication over an AWGN channel.}
    \label{fig:AWGN_channel}
  \end{center}
\end{figure}
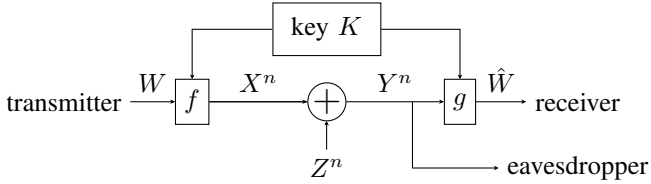

We consider the model illustrated in Fig.~\ref{fig:AWGN_channel}, in which a sender attempts to transmit a uniformly distributed message 
$W\in\mathcal{M}$ to a legitimate receiver over an AWGN channel in the presence of an eavesdropper, by encoding the message as a length $n$ codeword $X^n$. We assume that the eavesdropper and the legitimate receiver observe the same outputs 
\begin{IEEEeqnarray}{rCl}
  \label{eq:channel_awgn}
  Y_i=X_i+Z_i, \quad i=1,2,\ldots,n.
\end{IEEEeqnarray}
where $Z_i\sim\mathcal{N}(0,\sigma)$. The transmission is assisted by a uniformly distributed key $K\in\mathcal{K}$,
shared between the transmitter and receiver but unknown to the eavesdropper. Formally, a code $\mathcal{C}=(f,g)$ of length $n$ for message set $\mathcal{M}$ and key set $\mathcal{K}$ consists of an encoder $f\colon\mathcal{M} \times \mathcal{K} \rightarrow \mathbb{R}^n , (w,k) \mapsto x^n$ and a decoder $g\colon \mathbb{R}^n \times \mathcal{K} \rightarrow \mathcal{M}, (y^n,k) \mapsto \hat{w}$.  
The code $\mathcal{C}$ is publicly known to all parties. We denote a uniformly distributed input over the codewords of $\mathcal{C}$ by $X_{\mathcal{C}}^n$.
The reliability of the communication is measured in terms of the average probability of decoding error
\begin{IEEEeqnarray}{rCl}
  \label{average_probability_decoding_error_section_finite_blocklength}
  P_e^{\textnormal{avg}}&=&\frac{1}{|\mathcal{K}|}\frac{1}{|\mathcal{M}|}\sum_{k=1}^{|\mathcal{K}|}\sum_{w=1}^{|\mathcal{M}|}\mathbb{P}\left[g(Y^n, k) \neq w ~|~ x^n = f(w,k) \right].\nonumber\\
\end{IEEEeqnarray}
Covertness is measured in terms of the KL divergence ${D}\left(P_{Y^n_{\mathcal{C}}}\|P_{Z^n}\right)$ between the output distribution
\begin{IEEEeqnarray}{rCl}
  \label{output_sequence_averaged}
  P_{Y^n_\mathcal{C}}(\cdot)=\frac{1}{|\mathcal{K}||\mathcal{M}|}\sum_{k=1}^{|\mathcal{K}|}
  \sum_{w=1}^{|\mathcal{M}|}
  P_{Y^n|X^n} (\cdot|f(w,k))
\end{IEEEeqnarray}
induced by the coding scheme, and the output distribution $P_{Z^n}\sim\mathcal{N}(0,\sigma)$ induced in the absence of transmission. A code $\mathcal{C}$ of blocklength $n$, message size $M=\left|\mathcal{M}\right|$, that satisfies $P_e^{\textnormal{avg}}\leq \epsilon$ and 
\begin{IEEEeqnarray}{rCl}
\label{covertness_constraint}
{D}\left(P_{Y^n_{\mathcal{C}}}\|P_{Z^n}\right) \leq \delta
\end{IEEEeqnarray}
for some $\epsilon, \delta \in (0,1]$ is called an 
$(n, M, \epsilon, \delta)$-code.
We denote 
\begin{IEEEeqnarray}{rCl}
  M^*\left(n,\epsilon,\delta\right)&=&\max\bigg\{M~|~ \exists  (n,M,\epsilon,\delta)\textnormal{-code} \bigg\}.
\end{IEEEeqnarray}
The covert capacity is then defined as 
\begin{IEEEeqnarray}{rCl}
  L_{\epsilon,\delta}=\liminf\limits_{n\rightarrow +\infty} \frac{\ln\left(M^*(n,\epsilon,\delta)\right)}{\sqrt{n}}.
\end{IEEEeqnarray}
In what follows, we do not characterize the number of key nats required and merely assume it is large enough, in a way we make more precise later.

Our main results are summarized in the following propositions.
\begin{proposition}
  \label{prop:converse}
  For the AWGN channel defined by \eqref{eq:channel_awgn} 
  \begin{IEEEeqnarray}{rCl}
    \label{lower_bound_m_star_kl}
    L_{\epsilon,\delta}&\leq&\frac{\sqrt{\delta}}{1-\epsilon}. 
  \end{IEEEeqnarray}
\end{proposition}

\begin{proposition}
  \label{prop:achievability}
  For the AWGN channel defined by \eqref{eq:channel_awgn}
  \begin{IEEEeqnarray}{rCl}
    \label{upper_bound_m_star_kl}
    L_{\epsilon,\delta}&\geq&\frac{\sqrt{\delta}}{\sqrt{1-\epsilon}}.
  \end{IEEEeqnarray}
\end{proposition}

\begin{figure}[!htbp]
  \begin{center}
    \includegraphics[width=0.5\textwidth]{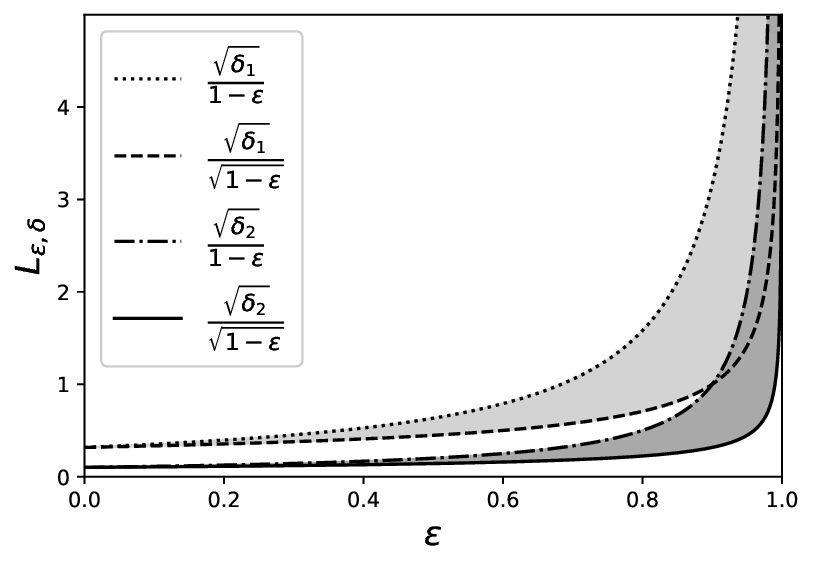}
    \caption{Upper and lower bounds on  $L_{\epsilon,\delta}$ for 
      $\delta_1=0.1$ and $\delta_2=0.01$.}
    \label{fig:comparison_lower_upper_bound_L_epsilon}
  \end{center}
\end{figure}

Although the bounds are not tight, in the regime $\epsilon,\delta \ll 1$, we obtain
\begin{equation}
  \sqrt{\delta}\left(1+\frac{\epsilon}{2}\right) \lesssim L_{\epsilon,\delta}\lesssim \sqrt{\delta}(1+\epsilon).
\end{equation}
Propositions~\ref{prop:converse} and~\ref{prop:achievability} are illustrated in Fig.~\ref{fig:comparison_lower_upper_bound_L_epsilon}, showing a relatively close characterization of $L_{\epsilon,\delta}$ over a wide range of average error probability $\epsilon$. 

Most importantly, our results clearly show that $L_{\epsilon,\delta}>\lim\limits_{\epsilon'\rightarrow 0} L_{\epsilon',\delta}$ for $\epsilon>0$.

\section{Proof of Proposition~\ref{prop:converse}}
\label{sec:converse}

Consider an $(n, M, \epsilon,\delta)$-code, that is $P_e^{\textnormal{avg}}\leq \epsilon$ and ${D}\left(P_{Y^n_{\mathcal{C}}}\|P_{Z^n}\right) \leq \delta$. 
Let $P_{\bar{X}}$ denote the average input distribution over the secret key, the uniform message and the $n$ channel uses, and $P_{\bar{Y}}$ the corresponding output distribution.
The average power of the code is 
\begin{IEEEeqnarray}{rCl}
  \rho_n=\mathbb{E}[\bar{Y}^2]=\frac{1}{n} \frac{1}{\abs{\mathcal{K}}\abs{\mathcal{M}}} \sum_{k=1}^{\abs{\mathcal{K}}} \sum_{w=1}^{\abs{\mathcal{M}}} \norm{f(w,k)}_2^2.
\end{IEEEeqnarray}
We know from~\cite[equation (75)]{fundamental_covertness} that the covertness constraint implies the average power constraint 
\begin{IEEEeqnarray}{rCl}
\label{power_constraint}
  \rho_n \leq 2\sigma^2 \sqrt{\frac{\delta}{n}}+O\left(\frac{1}{n}\right).
\end{IEEEeqnarray}
Following \cite[Section III]{fundamental_covertness}, for each realization $\mathsf{K}=k$ of the secret key, Fano's inequality implies
    \begin{IEEEeqnarray}{rCl}
     \ln\left|\mathcal{M}\right|(1-\epsilon_k)- 1
    &\leq& I(X^n_{\mathcal{C}_k};Y^n_{\mathcal{C}_k})
  \end{IEEEeqnarray}
  where $\epsilon_k$ is the average error probability when $\mathsf{K}=k$, $\mathcal{C}_k$ is the subcode for key $k$, $P_{X^n_{\mathcal{C}_k}}$ is the uniform distribution over $\mathcal{C}_k$ and $P_{Y^n_{\mathcal{C}_k}}=1/|\mathcal{M}| ~
  \sum_{w=1}^{|\mathcal{M}|}
  P_{Y^n|X^n} (\cdot|f(w,k))$ is the corresponding output distribution. By averaging over the key,
  we obtain
\begin{IEEEeqnarray}{rCl}
    \ln\left|\mathcal{M}\right|(1-\epsilon)- 1 \nonumber 
    &\leq& \frac{1}{|\mathcal{K}|} \sum_{k=1}^{|\mathcal{K}|}I(X^n_{\mathcal{C}_k};Y^n_{\mathcal{C}_k})\nonumber\\
    \label{markov_chain_inequality}
    &\leq&I(X^n_\mathcal{C};Y^n_\mathcal{C})\\ 
    \label{use_convexity_mutual_information}
  &\leq& nI(\bar{X};\bar{Y}) \\
  \label{gaussian_maximizes_entropy}
  & \leq& \frac{n}{2} \ln \left(1+\frac{\rho_n}{\sigma^2}\right)\\
  &=&\sqrt{\delta}\sqrt{n}+O\left(1\right)
\end{IEEEeqnarray}
where \eqref{markov_chain_inequality} follows by the concavity of the mutual information with respect to the input distribution; \eqref{use_convexity_mutual_information} by the chain rule and the concavity of the mutual information with respect to the input distribution; \eqref{gaussian_maximizes_entropy} 
since the Gaussian distribution maximizes $I(\bar{X};\bar{Y})$ among all distributions with second moment $\rho_n$. Taking the limit as $n\rightarrow\infty$ concludes the proof.

\section{Proof of Proposition~\ref{prop:achievability}}
\label{sec:achievability}


In the following, we establish a lower bound for $\ln(M^*\left(n,\epsilon,\delta\right))$. The achievability proof 
follows the ideas of \cite[Appendix A]{tahmasbi2018first} and \cite[Theorem 77]{polyanskiythesis}, and
consists of two steps. 
First, we prove an intermediate Lemma.
\begin{lemma} \label{lemma_first_order}
 Let $\epsilon'_n=n^{-1/6}$, and $\delta' >0$.  
Over the AWGN channel \eqref{eq:channel_awgn},
there exists an 
$(n,M,\epsilon'_n,\delta')$-code $\mathcal{C}'$
such that 
\begin{IEEEeqnarray}{rCl}
\label{lower_bound_ln_M_L_epsilon}
\ln M \geq \sqrt{\delta'}\sqrt{n}+O(n^{1/3}).
 \end{IEEEeqnarray}
 \end{lemma}
 The proof of Lemma \ref{lemma_first_order} is based on a random coding construction using BPSK inputs, and is provided in Appendix \ref{Appendix_proof_Lemma_first_order}.
Given $\delta>0$ and $\epsilon>0$, define 
\begin{IEEEeqnarray}{rCl}
\label{delta_prime_lower_bound_L_epsilon}
     \delta'=\frac{1-\epsilon'_n}{1-\epsilon}\delta,
\end{IEEEeqnarray}
where $\epsilon'_n=n^{-1/6}$. Lemma 1 guarantees the existence of an $(n,M,\epsilon_n',\delta')$ code $\mathcal{C}'$ whose size is lower bounded by \eqref{lower_bound_ln_M_L_epsilon}. 
We now extend $\mathcal{C}'$ by adding a suitable number of all-zero codewords, in order to obtain an $(n, M,\epsilon,\delta)$-code of the required size.  
Specifically, we construct a new code $\mathcal{C}$ of size $M \frac{1-\epsilon'_n}{1-\epsilon}$ by adding $\alpha_n M$ all-zero codewords (independently of the value of the key), where
\begin{IEEEeqnarray}{rCl}
    \alpha_n= \frac{\epsilon-\epsilon'_n}{1-\epsilon}.
\end{IEEEeqnarray} 
Indeed, given than $\epsilon>\epsilon'_n=n^{-1/6}$ for sufficiently large $n$, the average error probability $P_e^{\text{avg}}$ of this new code admits $\epsilon$ as an upper bound:
\begin{IEEEeqnarray}{rCl}
    P_e^{\text{avg}}&\leq&\frac{1}{1+\alpha_n}\epsilon'_n+\frac{\alpha_n}{1+\alpha_n}~=~\epsilon.
\end{IEEEeqnarray}
Furthermore by convexity of the Kullback-Leibler divergence,
\begin{IEEEeqnarray}{rCl}
\IEEEeqnarraymulticol{3}{l}{
    D(P_{Y^n_\mathcal{C}}\|P_{Z^n})
    }\nonumber\\* \quad
&\leq& \frac{1}{1+\alpha_n} D(P_{Y^n_{\mathcal{C}'}}\|P_{Z^n}) + \frac{\alpha_n}{1+\alpha_n} D(P_{Z^n}\|P_{Z^n})\nonumber\\
&=&\frac{1}{1+\alpha_n}\delta'\nonumber\\
    &=&\frac{1-\epsilon}{1-\epsilon'_n}\delta'\nonumber\\
    &=&\delta.
\end{IEEEeqnarray}
Thus, $\mathcal{C}$ is an $(n,M\frac{1-\epsilon'_n}{1-\epsilon},\epsilon,\delta)$-code such that
\begin{IEEEeqnarray}{rCl}
\IEEEeqnarraymulticol{3}{l}{
    \ln\left(M\frac{1-\epsilon'_n}{1-\epsilon}\right)
    }\nonumber\\* \quad
&\geq&\sqrt{\delta'}\sqrt{n}+O(n^{1/3})+\ln\left(\frac{1-\epsilon'_n}{1-\epsilon}\right)\nonumber\\
\label{use_definition_delta_prime}
&=& \sqrt{\frac{1-\epsilon'_n}{1-\epsilon}}\sqrt{\delta}\sqrt{n}+O\left(n^{1/3}\right)+\ln\left(\frac{1-\epsilon'_n}{1-\epsilon}\right)\\
\label{use_choice_epsilon_prime_lower_bound_L_epsilon} 
&=& \sqrt{\frac{\delta}{1-\epsilon}}\sqrt{n}+O\left(n^{1/3}\right),
\end{IEEEeqnarray}
where \eqref{use_definition_delta_prime}
follows by \eqref{delta_prime_lower_bound_L_epsilon}, 
and \eqref{use_choice_epsilon_prime_lower_bound_L_epsilon} since $\epsilon_n'=n^{-1/6}$.  This concludes the proof of Proposition \ref{prop:achievability}. 






\appendix
\subsection{Proof of Lemma \ref{lemma_first_order}}
\label{Appendix_proof_Lemma_first_order}
 We denote
 \begin{IEEEeqnarray}{rCl}
 \label{value_P_finite_blocklength_bpsk}
 P=2\sigma^2\sqrt{\frac{\delta'}{n}},
 \end{IEEEeqnarray}
and 
consider the uniform distribution $P_X$ on $\{-\sqrt{P},\sqrt{P}\}$. Let $P_Y$ be the corresponding output distribution through $P_{Y|X}$: 
\begin{IEEEeqnarray}{rCl}
p_{Y}(y)=\frac{1}{2}\frac{1}{\sqrt{2\pi}\sigma}e^{-\frac{(y-\sqrt{P})^2}{2\sigma^2}}+\frac{1}{2}\frac{1}{\sqrt{2\pi}\sigma}e^{-\frac{(y+\sqrt{P})^2}{2\sigma^2}}.
\end{IEEEeqnarray}
We use the notation $P_{X^n}=P_X^{\otimes n}$ and $P_{Y^n}=P_Y^{\otimes n}$ for the corresponding i.i.d. product distributions.
We generate a random code $\mathsf{C}$ by picking every codeword i.i.d. from $P_X^{\otimes n}$. We denote by $X_{w,k}^n$ for $w=1,\ldots,|\mathcal{M}|,k=1,\ldots,|\mathcal{K}|$ the random codewords, by $X_{w,k,i}$ the $i$\textsuperscript{th} component of $X_{w,k}^n$, and by $P_{Y^n_{\mathsf{C}}}$ the output distribution of the random code.
We want to show that this code satisfies the covertness constraint $\delta'$ with high probability. First, we notice that
\begin{IEEEeqnarray}{rCl}
\IEEEeqnarraymulticol{3}{l}{\mathbb{P} \left[D\left(P_{Y^n_{\mathsf{C}}}\|P_{Z^n}\right) > \delta' \right]
} \nonumber \\*
\label{probability_to_be_covert}
&=& \mathbb{P} \left[D\left(P_{Y^n_{\mathsf{C}}}\|P_{Y^n}\right) + \mathbb{E}_{P_{Y^n_{\mathsf{C}}}} \left[ \ln\left(\frac{p_{Y^n}(Y^n_{\mathsf{C}})}{p_{Z^n}(Y^n_{\mathsf{C}})}\right) \right] > \delta' \right].~~~~~~~
\end{IEEEeqnarray}
We compute
\begin{IEEEeqnarray}{rCl}
\IEEEeqnarraymulticol{3}{l}{
\mathbb{E}_{P_{Y^n_{\mathsf{C}}}} \left[ \ln\left(\frac{p_{Y^n}(Y^n_{\mathsf{C}})}{p_{Z^n}(Y^n_{\mathsf{C}})}\right) \right]
} \nonumber \\*
&=& \int_{\mathbb{R}^n} p_{Y^n_{\mathsf{C}}}(y^n) \ln\left(
\prod_{i=1}^n \frac{\frac{1}{2} \left( e^{-\frac{(y_i - \sqrt{P})^2}{2\sigma^2}} + e^{-\frac{(y_i + \sqrt{P})^2}{2\sigma^2}}\right)}{e^{-\frac{y_i^2}{2}}}\right) \dd y^n \nonumber\\
&=&  \int_{\mathbb{R}^n} p_{Y^n_{\mathsf{C}}}(y^n) \left( \sum_{i=1}^n \ln \left(
\frac{e^{\frac{y_i \sqrt{P}}{\sigma^2}} + e^{-\frac{y_i \sqrt{P}}{\sigma^2}}}{2}\right)  - \frac{nP}{2\sigma^2} \right) \dd y^n \nonumber\\
&\stackrel{(a)}{\leq}&  \int_{\mathbb{R}^n} p_{Y^n_{\mathsf{C}}}(y^n) \sum_{i=1}^n \left(
\frac{y_i^2P}{2\sigma^4} - \frac{y_i^4P^2}{12\sigma^8} + \frac{y_i^6P^3}{45\sigma^{12}} \right) - \frac{nP}{2\sigma^2} \dd y^n \nonumber \\
&=& \frac{1}{|\mathcal{K}|\cdot |\mathcal{M}|} \sum_{k=1}^{|\mathcal{K}|} \sum_{w=1}^{|\mathcal{M}|}  \int_{\mathbb{R}^n} \prod_{i=1}^n \frac{e^{-\frac{(y_i-X_{w,k,i})^2}{2}}}{\sqrt{2\pi}\sigma}  \nonumber \\  
&&\cdot \sum_{i=1}^n \left[
\frac{y_i^2P}{2\sigma^4} -\frac{y_i^4P^2}{12\sigma^8} + \frac{y_i^6P^3}{45\sigma^{12}}
    \right]\dd y^n - \frac{nP}{2\sigma^2} \nonumber\\
&\stackrel{(b)}{=}& 
    n\Bigg(
\left(\sigma^2 + P\right) \frac{P}{2\sigma^4}  -\left(3\sigma^4 + 6\sigma^2 P  + P^2\right) \frac{P^2}{12\sigma^8} \nonumber \\
\label{expectation_covertness}
&&+\left(15\sigma^6 + 45\sigma^4 P 
+ 15\sigma^2 P^2 + P^3\right) \frac{P^3}{45\sigma^{12}}
    \Bigg) - \frac{nP}{2\sigma^2},
\end{IEEEeqnarray}%
where 
$(a)$ follows by the inequality $\ln\left(\cosh(x)\right)\leq \frac{x^2}{2}-\frac{x^4}{12}+\frac{x^6}{45}$,
and (b) holds because $\mathbb{P}\left[X_{w,k,i}^2=P\right]=1$ for any $w,k,i$.
Thus, from \eqref{probability_to_be_covert} and \eqref{expectation_covertness}
\begin{IEEEeqnarray}{rCl}
\IEEEeqnarraymulticol{3}{l}{
\mathbb{P} \left[D\left(P_{Y^n_{\mathsf{C}}}\|P_{Z^n}\right) > \delta' \right]
} \nonumber \\*
&=& \mathbb{P} \Bigg[D\left(P_{Y^n_{\mathsf{C}}}\|P_{Y^n}\right) +n \Bigg(\frac{P^2}{4\sigma^4}-\frac{P^3}{6\sigma^6}+\frac{11}{12}\frac{P^4}{\sigma^8} \nonumber \\
&&~~{}+\frac{P^5}{3\sigma^{10}}+\frac{P^6}{45\sigma^{12}}\Bigg)>\delta' \Bigg] \nonumber \\
&\stackrel{(a)}{=}&\mathbb{P} \left[D\left(P_{Y^n_{\mathsf{C}}}\|P_{Y^n}\right) >\frac{nP^3}{6\sigma^6} - \frac{11}{12}\frac{nP^4}{\sigma^8} - \frac{nP^5}{3\sigma^{10}} - \frac{nP^6}{45\sigma^{12}} \right] \nonumber\\
&\stackrel{(b)}{\leq}&\frac{\mathbb{E}\left[D\left(P_{Y^n_{\mathsf{C}}}\|P_{Y^n}\right) \right]}{\frac{4(\delta')^\frac{3}{2}}{3\sqrt{n}} + O\left(\frac{1}{n}\right)} \nonumber\\
\label{check_covertness_bpsk_hypothesis_awgn}
&\stackrel{(c)}{=}&O\left(\sqrt{n}e^{-n}\right),
\end{IEEEeqnarray}
where $(a)$ holds 
by \eqref{value_P_finite_blocklength_bpsk}; 
$(b)$ follows by Markov's inequality and \eqref{value_P_finite_blocklength_bpsk}; and $(c)$ 
by
Hayashi and Matsumoto's channel resolvability bound \cite[Theorem 14]{hayashi2016secure}  which ensures that choosing a suitably fitted key size $\ln|\mathcal{K}|=O(n)$ yields \eqref{check_covertness_bpsk_hypothesis_awgn} (see \cite[Proposition 1]{bouette_journal_covert_over_additive_channels}).

We now prove the existence of an $(n,M,\epsilon'_n,\delta')$ code. 
From Shannon's achievability bound \cite[Theorem 1]{shannon1957certain}\cite[Theorem 18.5]{info_theory_polyanskiy}, we know that for all $\gamma>0$,
\begin{IEEEeqnarray}{rCl}
\IEEEeqnarraymulticol{3}{l}{
\mathbb{E}_{\mathsf{C}}[P_e^{\text{avg}}(\mathsf{C})]
} \nonumber \\* ~
\label{achievability_finite_blocklength_bpsk}
    &\leq& \mathbb{P}\left[i_{X^n,Y^n}(X^n,Y^n) \leq \ln\left|\mathcal{M}\right| + n\gamma \right] + \exp(-n\gamma),
\end{IEEEeqnarray}
where $P_e^{\text{avg}}(\mathsf{C})$ is the average error probability of the random code $\mathsf{C}$, and
\begin{IEEEeqnarray}{rCl}
i_{X^n,Y^n}(x^n,y^n)&=&\ln\left(\frac{p_{Y^n|X^n}(y^n|x^n)}{p_{Y^n}(y^n)}\right), \quad \forall x^n, y^n \in \mathbb{R}^n\nonumber\\    
\end{IEEEeqnarray}
is the information density of the joint distribution $P_{X^n,Y^n}$.
We now assume that $\ln\left|\mathcal{M}\right|$ is chosen so that
\begin{IEEEeqnarray}{rCl}
\label{condition_mean_upper_bound_half_epsilon}
    \mathbb{E}_{\mathsf{C}}[P_e^{\text{avg}}(\mathsf{C})]\leq\frac{\epsilon'_n}{2},
\end{IEEEeqnarray}
where $\epsilon'_n=n^{-1/6}$. 
A sufficient condition for this will be derived later. Then, by Markov's inequality, 
\begin{IEEEeqnarray}{rCl}
\label{use_markov_2_epsilon_achievability_average_error_probability}
    \mathbb{P}\left[P_e^{\text{avg}}(\mathsf{C})\leq\epsilon'_n\right]
    &\geq&\frac{1}{2}
\end{IEEEeqnarray}
and it follows that there exists a code $\mathcal{C}$ with average error probability $P_e^{\text{avg}} \leq \epsilon'_n$ which satisfies the covertness constraint $\delta'$, since 
\begin{IEEEeqnarray}{rCl}
\IEEEeqnarraymulticol{3}{l}{
\mathbb{P}\left[\left\{P_e^{\text{avg}}(\mathsf{C})\leq\epsilon'_n\right\}\cap \left\{D\left(P_{Y^n_{\mathsf{C}}}\|P_{Z^n}\right)\leq\delta'\right\}\right]
} \nonumber \\* \qquad
&\geq& 1-\mathbb{P}\left[P_e^{\text{avg}}(\mathsf{C})>\epsilon'_n\right]-\mathbb{P}\left[D\left(P_{Y^n_{\mathsf{C}}}\|P_{Z^n}\right)>\delta'\right]\nonumber\\
    &\geq& \frac{1}{2}-\mathbb{P}\left[D\left(P_{Y^n_{\mathsf{C}}}\|P_{Z^n}\right)>\delta'\right]\nonumber\\    \label{use_check_covertness_bpsk_hypothesis_awgn}
    &>& 0,
\end{IEEEeqnarray}
where \eqref{use_check_covertness_bpsk_hypothesis_awgn} follows by \eqref{check_covertness_bpsk_hypothesis_awgn} for large enough $n$.

We now find a sufficient condition for \eqref{condition_mean_upper_bound_half_epsilon} to hold. 
Note that for any $x^n,y^n\in\mathbb{R}^n$, 
\begin{IEEEeqnarray}{rCl}
    \IEEEeqnarraymulticol{3}{l}{i_{X^n,Y^n}(x^n,y^n)}\nonumber \\* ~~~~ 
    &=&\ln\left(\frac{\prod_{i=1}^n\frac{1}{\sqrt{2\pi}\sigma} e^{-\frac{(y_i-x_i)^2}{2\sigma^2}}}{\prod_{i=1}^n\frac{1}{\sqrt{2\pi}\sigma} \frac{1}{2} \left(e^{-\frac{(y_i-\sqrt{P})^2}{2\sigma^2}}+e^{-\frac{(y_i+\sqrt{P})^2}{2\sigma^2}}\right)}\right)\nonumber\\
    &=&\sum_{i=1}^n\left(\ln\left(\frac{e^{\frac{x_iy_i}{\sigma^2}}}{ \frac{1}{2} \left(e^{\frac{\sqrt{P}y_i}{\sigma^2}}+e^{-\frac{\sqrt{P}y_i}{\sigma^2}}\right)}\right)+\frac{P-x_i^2}{2\sigma^2}\right)\nonumber\\
    &=&\sum_{i=1}^n\left(\frac{x_iy_i}{\sigma^2}- \ln\left(\cosh\left(\frac{\sqrt{P}y_i}{\sigma^2}\right)\right)+\frac{P-x_i^2}{2\sigma^2}\right)\nonumber\\
    \label{second_use_dl_cosh}
    &\geq&\sum_{i=1}^n\left(\frac{x_iy_i}{\sigma^2}- \frac{P y_i^2}{2\sigma^4}+\frac{P-x_i^2}{2\sigma^2}\right),
\end{IEEEeqnarray}
where \eqref{second_use_dl_cosh} follows by the inequality $\ln\left(\cosh(x)\right)\leq \frac{x^2}{2}$.
Therefore 
\begin{IEEEeqnarray}{rCl}
\IEEEeqnarraymulticol{3}{l}{
\mathbb{P}\left[i_{X^n,Y^n}(X^n,Y^n) \leq \ln\left|\mathcal{M}\right| + n\gamma \right]
}\nonumber \\* ~
&\stackrel{(a)}{\leq}&  \mathbb{P}\left[\sum_{i=1}^n\left(\frac{X_iY_i}{\sigma^2}- \frac{P Y_i^2}{2\sigma^4}\right) \leq \ln\left|\mathcal{M}\right| + n\gamma \right] \nonumber \\
    &\stackrel{(b)}{=}&\mathbb{P}\Bigg[\sum_{i=1}^n\left(\frac{X_iZ_i}{\sigma^2}- \frac{P (Z_i^2+2X_iZ_i)}{2\sigma^4}\right)+\frac{nP}{\sigma^2}-\frac{nP^2}{2\sigma^4} \nonumber \\ 
    &&{}~~~\leq \ln\left|\mathcal{M}\right| + n\gamma \Bigg]\nonumber\\
\label{information_density_finite_blocklength_bpsk}
    &=&\mathbb{P}\Bigg[\sum_{i=1}^n \left(\left(\frac{1}{\sigma^2}-\frac{P}{\sigma^4}\right)X_iZ_i-\frac{P}{2\sigma^4}Z_i^2\right) +\frac{nP}{\sigma^2}-\frac{nP^2}{2\sigma^4} \nonumber\\
  &&{}~~~ \leq  \ln\left|\mathcal{M}\right| + n\gamma \Bigg]
\end{IEEEeqnarray}
where 
$(a)$ and $(b)$ follow because $\mathbb{P}\left[X_i^2=P\right]=1$. 
For simplicity, we denote the random expression inside \eqref{information_density_finite_blocklength_bpsk} by
\begin{IEEEeqnarray}{rCl}
\label{def_wi_hypothesis_finite_blocklength_awgn_bpsk_achievability}
U_i=  \left(\frac{1}{\sigma^2}-\frac{P}{\sigma^4}\right)X_iZ_i-\frac{P}{2\sigma^4}Z_i^2 \quad \forall i=1,\ldots,n.
\end{IEEEeqnarray}
Note that for $i=1,\dots,n$, $U_i$ is equal to
\begin{IEEEeqnarray}{rCl}
    U_i'=  \left(\frac{1}{\sigma^2}-\frac{P}{\sigma^4}\right)\sqrt{P}Z_i-\frac{P}{2\sigma^4}Z_i^2 
\end{IEEEeqnarray}
with probability $1/2$ and to
\begin{IEEEeqnarray}{rCl}
    U_i''=  -\left(\frac{1}{\sigma^2}-\frac{P}{\sigma^4}\right)\sqrt{P}Z_i-\frac{P}{2\sigma^4}Z_i^2 
\end{IEEEeqnarray}
with probability $1/2$. Furthermore $Z_i\sim\mathcal{N}(0,\sigma^2)$ follows the same distribution as $-Z_i$, so $U_i'\sim U_i''$. Therefore we can restrict ourselves to the case $U_i=U_i'$. For all $i=1,\dots,n$, we have 
    $\mathbb{E}[U_i]=-\frac{P}{2\sigma^2}$
and
\begin{IEEEeqnarray}{rCl}
\IEEEeqnarraymulticol{3}{l}{
\mathrm{Var}\left[U_i\right]
}\nonumber \\* \qquad
    &=&\mathbb{E}\Bigg[\left(\frac{1}{\sigma^2}-\frac{P}{\sigma^4}\right)^2PZ_i^2+2\left(\frac{1}{\sigma^2}-\frac{P}{\sigma^4}\right)\frac{P\sqrt{P}}{2\sigma^4}Z_i^3 \nonumber\\
    &&{}~~~+\frac{P^2}{4\sigma^8}Z_i^4\Bigg] -\frac{P^2}{4\sigma^4}\nonumber\\
    &=&\sigma^2 P\left(\frac{1}{\sigma^2}-\frac{P}{\sigma^4}\right)^2 +3\sigma^4\frac{P^2}{4\sigma^8}-\frac{P^2}{4\sigma^4}\nonumber\\
    &=&\frac{P}{\sigma^2}- \frac{3 P^2}{2\sigma^4}+\frac{P^3}{\sigma^6}.\label{V_bpsk_hypothesis_awgn}
\end{IEEEeqnarray}
Injecting \eqref{information_density_finite_blocklength_bpsk} in \eqref{achievability_finite_blocklength_bpsk}, we obtain
\begin{IEEEeqnarray}{rCl}
\IEEEeqnarraymulticol{3}{l}{
\mathbb{E}_{\mathsf{C}}[P_e^{\text{avg}}(\mathsf{C})]
}\nonumber \\*
\label{before_use_berry_esseen_finite_blocklength_bpsk_proof_theorem_lower_bound_L_epsilon}
&\leq&\mathbb{P}\left[\sum_{i=1}^n \left(U_i-\mathbb{E}\left[U_i\right]\right) \leq \ln\left|\mathcal{M}\right|+n\gamma - \frac{nP}{2\sigma^2} +\frac{nP^2}{2\sigma^4} \right] \nonumber \\
&&{}+e^{-n\gamma}.  
\end{IEEEeqnarray}%
We want to show that
\eqref{before_use_berry_esseen_finite_blocklength_bpsk_proof_theorem_lower_bound_L_epsilon} can be upper bounded by $\frac{\epsilon_n'}{2}=\frac{n^{-1/6}}{2}$ when choosing
 \begin{IEEEeqnarray}{rCl}
 \ln\left|\mathcal{M}\right| = - n\gamma + \frac{nP}{2\sigma^2} - \frac{nP^2}{2\sigma^4}-n^\frac{1}{3},
 \end{IEEEeqnarray}
 with
     $\gamma=\frac{\ln(4n^\frac{1}{6})}{n}$,
 which satisfies \eqref{lower_bound_ln_M_L_epsilon}.
First, we prove that
\begin{IEEEeqnarray}{rCl}
\label{probability_error_small_exists_decomposition_bpsk}
\mathbb{P}\left[\sum_{i=1}^n \left(U_i-\mathbb{E}\left[U_i\right]\right) \leq -n^\frac{1}{3} \right]&\leq&\frac{1}{4}n^{-\frac{1}{6}}.
\end{IEEEeqnarray}
This follows from a result in large deviation theory \cite[Theorem 3.7.1]{large_deviation}. The statement of the theorem and further details are provided in Appendix \ref{appendix_moderate_deviations}. 
Applying \cite[Theorem 3.7.1]{large_deviation} with $\Gamma=(-\infty,-1]$, $a_n=n^{-\frac{1}{6}}$ and $V_i=n^{\frac{1}{4}}(U_i-\mathbb{E}[U_i])$, we obtain
\begin{IEEEeqnarray}{rCl}
    \mathbb{P}\left[\sum_{i=1}^n \left(U_i -\mathbb{E}[U_i]\right) <- n^\frac{1}{3}\right]
    &=&\mathbb{P}\left[n^{-\frac{7}{12}}\sum_{i=1}^n V_i <- 1\right]\nonumber\\
    &\leq&e^{-\frac{1}{2\mathrm{Var}[V_i]n^{-\frac{1}{6}}}+o\left(n^\frac{1}{6}\right)}\nonumber\\
    &\stackrel{(a)}{=}&e^{-\frac{\sqrt{1-\epsilon}}{4\sqrt{1-\epsilon'_n}\sqrt{\delta}}n^\frac{1}{6}+o\left(n^\frac{1}{6}\right)}\nonumber\\
    &\leq&e^{-\frac{\sqrt{1-\epsilon}}{4\sqrt{\delta}}n^\frac{1}{6}+o\left(n^\frac{1}{6}\right)}\nonumber\\   \label{end_computation_error_proba_vanishing_n_1_6}
    &=&o(n^{-\frac{1}{6}}),
\end{IEEEeqnarray}
where $(a)$ 
follows by \eqref{V_bpsk_hypothesis_awgn}, 
the definition of $P$ in \eqref{value_P_finite_blocklength_bpsk}, and \eqref{delta_prime_lower_bound_L_epsilon}.
Note that \eqref{end_computation_error_proba_vanishing_n_1_6} implies \eqref{probability_error_small_exists_decomposition_bpsk}, then combining \eqref{probability_error_small_exists_decomposition_bpsk} and \eqref{before_use_berry_esseen_finite_blocklength_bpsk_proof_theorem_lower_bound_L_epsilon} ensures
\begin{IEEEeqnarray}{rCl}
\mathbb{E}\left[P_e^{\text{avg}}(\mathsf{C})\right]&\leq&\mathbb{P}\left[\sum_{i=1}^n \left(U_i-\mathbb{E}\left[U_i\right]\right) \leq -n^\frac{1}{3} \right]+e^{-n\gamma}\nonumber\\
&\leq&\frac{1}{2}n^{-\frac{1}{6}}=\frac{\epsilon'_n}{2}
\end{IEEEeqnarray}
which establishes \eqref{condition_mean_upper_bound_half_epsilon} as claimed. This concludes the proof of Lemma \ref{lemma_first_order}.  

\subsection{Moderate deviations bound}
\label{appendix_moderate_deviations}
We restate \cite[Theorem 3.7.1]{large_deviation}, which is used to prove \eqref{probability_error_small_exists_decomposition_bpsk}.
\begin{theorem}[Moderate Deviations]
\label{appendix_theorem_moderate_deviations}
Let $V_1,\ldots,V_n$ be a sequence of $\mathbb{R}$-valued i.i.d. random variables with variance $\sigma^2$ such that for any $i=1,\ldots,n$, $f(\lambda)=\ln(\mathbb{E}[e^{\lambda V_i}]) < \infty$ in some ball around the origin and $\mathbb{E}[V_i]=0$. Fix $a_n\rightarrow 0$ such that $n a_n \rightarrow+\infty$ as $n\rightarrow +\infty$, and let $A_n = \sqrt{\frac{a_n}{n}} \sum_{i=1}^n V_i$. Then, for every measurable set $\Gamma$,
\begin{IEEEeqnarray}{rCl}
    \limsup\limits_{n\rightarrow +\infty} a_n \ln\left(\mathbb{P}\left[A_n\in\Gamma\right]\right)&\leq&-\frac{1}{2}\inf_{x\in\bar{\Gamma}} \frac{x^2}{\sigma^2},
\end{IEEEeqnarray}
where $\bar{\Gamma}$ is the closure of the set $\Gamma$.
\end{theorem}
We now check that the hypotheses of the theorem are satisfied for $V_i=n^{\frac{1}{4}}(U_i -\mathbb{E}[U_i])$, so that \eqref{probability_error_small_exists_decomposition_bpsk} 
holds. 
We compute
\begin{IEEEeqnarray}{rCl}
    \IEEEeqnarraymulticol{3}{l}{
    \ln\left(\mathbb{E}\left[e^{\lambda n^\frac{1}{4}\left(U_i -\mathbb{E}[U_i]\right)}\right]\right)
    } \nonumber \\*
    \label{f_lambda}
    &=&\ln\left(\mathbb{E}\left[e^{\lambda \left(\left(\frac{1}{\sigma^2}-\frac{P}{\sigma^4}\right)n^\frac{1}{4}\sqrt{P}Z_i-n^\frac{1}{4}\frac{P}{2\sigma^4}Z_i^2+n^\frac{1}{4}\frac{P}{2\sigma^2}\right)}\right]\right)\nonumber\\
    &=&\ln\left(\mathbb{E}\left[e^{\lambda \left(-\frac{n^\frac{1}{4}P}{2\sigma^4}\left(Z_i-\left(\frac{\sigma^2}{P}-1\right)\sqrt{P}\right)^2+\frac{n^\frac{1}{4}}{2}-\frac{n^\frac{1}{4}P}{2\sigma^2}+\frac{n^\frac{1}{4}P^2}{2\sigma^4}\right)}\right]\right)\nonumber\\    
    &\stackrel{(a)}{=}&
        \frac{\lambda n^\frac{1}{4}}{2}\left(1-\frac{P}{\sigma^2}+\frac{P^2}{\sigma^4}\right)-\frac{1}{2}\ln\left(1+\frac{n^\frac{1}{4}P\lambda}{\sigma^2}\right) \nonumber\\
    &&{}-\frac{\frac{n^\frac{1}{4}P^2\lambda}{2\sigma^4} \left(\frac{\sigma^2}{P}-1\right)^2}{1+\frac{n^\frac{1}{4}P\lambda}{\sigma^2}} \nonumber\\
    &=&\frac{n^\frac{1}{4}P\lambda}{2\sigma^2}+\frac{\frac{\sqrt{n}P\lambda^2}{2\sigma^2}-\frac{\sqrt{n}P^2\lambda^2}{\sigma^4}+\frac{\sqrt{n}P^3\lambda^2}{2\sigma^6}}{1+\frac{n^\frac{1}{4}P\lambda}{\sigma^2}}\nonumber\\
    &&{}-\frac{1}{2}\ln\left(1+\frac{n^\frac{1}{4}P\lambda}{\sigma^2}\right) 
\end{IEEEeqnarray}
where $(a)$ 
can be established using the moment generating function of a non-central chi-squared random variable \cite[eq. (29.6)']{johnson1995continuous}.
Recalling the definition of $P$ in \eqref{value_P_finite_blocklength_bpsk}, we deduce that \eqref{f_lambda} is $<\infty$ in a ball around the origin  for any $n$, so that Theorem
\ref{appendix_theorem_moderate_deviations} 
holds.

\section*{Acknowledgment}
This work was supported in part by the French National Research Agency through 
CY Initiative of Excellence (grant Investissements d’Avenir ANR-16-IDEX-0008).

\IEEEtriggeratref{12}
\bibliographystyle{IEEEtran}

\end{document}